\documentclass[%
 reprint,
superscriptaddress,
twocolumn,
 amsmath,amssymb,
 aps,
]{revtex4-1}

\usepackage{graphicx}
\usepackage{dcolumn}
\usepackage{bm}
\usepackage{color}

\begin{document}

\title{Lifshitz Transition and Nontrivial H-Doping Effect in Cr-based Superconductor KCr$_3$As$_3$H$_x$}

\author{Si-Qi Wu}
 \affiliation{Department of Physics and Zhejiang Province Key Laboratory of Quantum Technology and Device, Zhejiang University, Hangzhou 310027, China}
\author{Chao Cao}
 \email[]{ccao@hznu.edu.cn}
 \affiliation{Condensed Matter Group, Department of Physics, Hangzhou Normal University, Hangzhou 311121, China}
\author{Guang-Han Cao}
 \email[]{ghcao@zju.edu.cn}
 \affiliation{Department of Physics and Zhejiang Province Key Laboratory of Quantum Technology and Device, Zhejiang University, Hangzhou 310027, China}
 \affiliation{State Key Lab of Silicon Materials, Zhejiang University, Hangzhou 310027, China}
 \affiliation{Collaborative Innovation Centre of Advanced Microstructures, Nanjing 210093, China}

\date{\today}

\begin{abstract}
We report the first-principles study on the H-intercalated Cr-based superconductor KCr$_3$As$_3$H$_x$. Our results show a paramagnetic ground state for KCr$_3$As$_3$H. The electronic structure consists of two quasi-one-dimensional (Q1D) Fermi-surfaces and one 3D Fermi-surface which are mainly contributed by Cr-d$_{z^2}$, d$_{x^2-y^2}$ and d$_{xy}$ orbitals. The bare electron susceptibility shows a $\Gamma$-centered imaginary peak, indicating possible ferromagnetic spin fluctuations. Upon moderate hole doping, the system undergoes a Lifshitz transition, which may enhance the Q1D feature of the system.
The Bader charge analysis and electron localization functions reveal a strong bonding nature of hydrogen in KCr$_3$As$_3$H, which results in a nontrivial electron doping in KCr$_3$As$_3$H.
\end{abstract}

\maketitle

\section{Introduction}
Since 2015, the discovery of the superconductivity in quasi-one-dimensional (Q1D) Cr-based compounds $A_2$Cr$_3$As$_3$ ($A$~=~K, \mbox{Rb}, Cs) has intrigued great research interests~\cite{K233.baojk,Rb233.tangzt,Cs233.tangzt,review.caogh}.
The crystal structures of these compounds consist of CrAs double-wall subnanotubes that are separated by $A^+$ cations. The large upper critical fields ($H_{c2}$) that exceed the Pauli paramagnetic limit and the nodal-line superconducting gap by penetration depth measurement revealed the unconventional nature of the superconductivity~\cite{18cgh.cpb,PeneDepth}.
The nuclear magnetic resonance (NMR) experiments demonstrated a power-law behavior of the nuclear spin-lattice relaxation rate, indicating a Tomonaga-Luttinger liquid (TLL) normal state~\cite{NMR.imai}. The absence of the Hebel-Slichter peak in the NMR measurements pointed to possible nodes in the superconducting gap functions of $A_2$Cr$_3$As$_3$~\cite{NMR.zgq}.
The density functional theory (DFT) calculations on K$_2$Cr$_3$As$_3$ depicted an electronic structure with two Q1D Fermi-surfaces (FS) and one 3D FS in the low energy region~\cite{K233.caoc,K233.hjp1}. The subsequent theoretical works based on 3- and 6-band Hubbard models proposed possible spin-triplet pairing symmetry in $A_2$Cr$_3$As$_3$~\cite{K233.hjp2,K233.hjp3,zhouy,djh}.

Shortly after the discovery of $A_2$Cr$_3$As$_3$, a new series of Q1D Cr-based compounds $A$Cr$_3$As$_3$ ($A$~=~K, Rb, Cs) was found by deintercalating half of the alkali metal atoms with ethanol~\cite{K133.baojk,133.tangzt}. The preliminary measurements show that the KCr$_3$As$_3$ has a spin-glass ground state, which is consistent with the DFT calculations and the relaxation behavior from the muon spin-rotation ($\mu$SR) experiments~\cite{K133.caoc,muSR133}.
However, in a later report on KCr$_3$As$_3$ single crystal samples, it was demonstrated that the KCr$_3$As$_3$ exhibits superconductivity at low temperature~\cite{K133.rza}.
Such a discrepancy remained puzzling until the neutron powder diffraction (NPD) works from Taddei \emph{et~al.}, which revealed the composition difference between the superconducting and non-superconducting samples~\cite{KHCr3As3_NPD}.
According to Taddei \emph{et~al}., there are H intercalations during the reaction between K$_2$Cr$_3$As$_3$ and ethanol. Hence the correct chemical formula for ``KCr$_3$As$_3$" should be KCr$_3$As$_3$H$_x$~\footnote{Different from Ref.~\cite{KHCr3As3_NPD}, the element order of the chemical formula has been adapted according to the Pauling electronegativity in this paper.}.
Due to the different synthesis procedures, there is a higher H content in the superconducting sample. It is the H content difference that causes the different ground state properties for superconducting and non-superconducting samples.

Given the presence of H in KCr$_3$As$_3$H$_x$, it is then natural to ask what is the effect of H doping on its electronic properties.
In Ref.~\cite{KHCr3As3_NPD}, preliminary DFT investigations have been carried out on KCr$_3$As$_3$H, which show that the presence of H induces effective electron doping. However, detailed properties such as electron susceptibilities and doping effects still remain to be investigated.
In this paper, we report our DFT calculation results on the magnetic ground state, electronic structure, and bare electron susceptibility of KCr$_3$As$_3$H. Our results demonstrate a paramagnetic ground state with possible ferromagnetic fluctuations in KCr$_3$As$_3$H. The band structure of KCr$_3$As$_3$H consists of three Fermi-surfaces contributed by Cr-d$_{z^2}$, d$_{x^2-y^2}$ and d$_{xy}$ orbitals. The effect of H doping is confirmed to lift the Fermi-level. Upon hole doping, the system undergoes a Lifshitz transition, which may induce an enhancement of the Q1D property. We also discussed the bonding nature of H atoms in KCr$_3$As$_3$H. The result reveals a strong bonding picture which takes the responsibility for the effective electron doping in KCr$_3$As$_3$H.

\section{Methods}
Our first-principle calculations were performed with density functional theory method, as implemented in the VASP package~\cite{VASP}. The Kohn-Sham wave functions were treated with projected augmented wave (PAW) method~\cite{PAW}. The core electrons of K-3s, K-3p, Cr-3p, and As-3d were treated as valence electrons~\footnote{In addition to the valence electrons of K-4s, Cr-3d, Cr-4s, As-4s and As-4p}. For exchange-correlation energy, we employed the Perdew, Burke and Enzerhoff-type functionals~\cite{GGA}. A 500 eV energy cutoff and a 6$\times$6$\times$12 $\Gamma$-centered $\bm{k}$-mesh were adopted after careful convergence tests.
The crystal structure we used for most calculations was obtained from the structural optimization. The initial structure was obtained by assuming full H-site occupation of the experimental structure~\cite{KHCr3As3_NPD}. The lattice parameters were fixed during the optimization.
To calculate the bare electron susceptibility of KCr$_3$As$_3$H, we constructed a tight-binding Hamiltionan with 50 atom-centered Wannier orbitals (30 Cr-d, 18 As-p, and 2 H-s)~\cite{wannier90}. The bare electron susceptibility $\chi_0$ was then calculated using:\\
\begin{widetext}
\begin{equation}
\label{Chi0}
\chi_0(\omega, \bm{q})~=~\frac{1}{N_{\bm{k}}}
\sum_{ww'}\sum_{mn\bm{k}}\frac{\left\langle w|m\bm{k}\right\rangle\left\langle m\bm{k}|w'\right\rangle\left\langle w'|n\bm{k}+\bm{q}\right\rangle\left\langle n\bm{k}+\bm{q}|w\right\rangle}{\omega+\epsilon_{n\bm{k}+\bm{q}}-\epsilon_{m\bm{k}}+i\eta^{+}}\times
[f(\epsilon_{m\bm{k}})-f(\epsilon_{n\bm{k}+\bm{q}})]
\end{equation}
\end{widetext}
Where $w$, $w'$ denote the Wannier orbitals; $m$, $n$ denote the band indices; $\epsilon_{m\bm{k}}$ represents the energy eigenvalue of band $m$ at reciprocal vector $\bm{k}$; and $f(\epsilon)$ represents the Fermi-Dirac distribution function at $\epsilon~+~\epsilon_F$. A $\bm{k}$-mesh of 96$\times$96$\times$197 was used during the susceptibility calculations, and the susceptibility at $\bm{q}$~=~$\Gamma$ was approximated with the value of $\bm{q}$~=~(0.001, 0.001, 0.001). The same tight-binding Hamiltionan was also used for Fermi-surface plotting. The doping effects were studied within rigid band approximation.

\section{Results}

\subsection{Magnetic Ground State}
\begin{figure}[h]
\includegraphics[width=8cm]{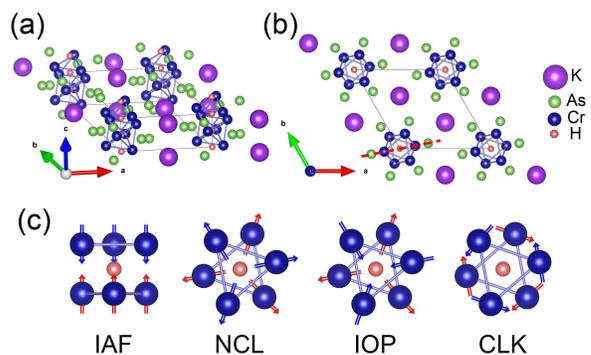}
\caption{\label{Struct}(a) Crystal structure of KCr$_3$As$_3$H. (b) Top view of the crystal structure. (c) Possible anti-ferromagnetic configurations of Cr octahedron, where red arrows denote the magnetic moments of the lower Cr triangle, blue arrows denote the moments of the upper Cr triangle.}
\end{figure}

We first investigated the ground state magnetism of KCr$_3$As$_3$H. The crystal structure of KCr$_3$As$_3$H is shown in Fig.~\ref{Struct}(a, b). Similar to KCr$_3$As$_3$, KCr$_3$As$_3$H holds a centrosymmetric structure (space group No.~176) where Cr and As atoms form double-wall subnanotubes along the $c$ axis. The nearest six Cr atoms, whose distances are 2.65~{\AA} and 2.60~{\AA} for intra- and inter-layer respectively, form an octahedron. The H atoms are located at the center of these octahedra. For possible anti-ferromagnetic (AFM) configurations, we considered the inter-layer collinear AFM (IAF), the noncollinear AFM (NCL), the noncollinear in-out co-planar AFM (IOP), and the noncollinear chiral-like AFM (CLK) types [Fig.~\ref{Struct}(c)]. In addition, we also considered the ferromagnetic (FM) and non-magnetic (NM) configurations.
In our present work, all the configurations are examined both with and without the spin-orbit coupling (SOC) effect.

\begin{table}
\caption{\label{tab:mag}Total energy differences and Cr magnetic moment for different magnetic configurations.}
\begin{ruledtabular}
\begin{tabular}{lcccc}
&\multicolumn{2}{c}{Not Relaxed}&\multicolumn{2}{c}{Relaxed{\footnotemark[1]}{\footnotemark[2]}}\\
&{${\Delta}E$ (meV/Cr)}&{$m{\rm _{Cr}}$ (${\mu}_B$)}&{${\Delta}E$ (meV/Cr)}&{$m{\rm _{Cr}}$ (${\mu}_B$)}\\
\hline
NM&0&---&0&---\\
FM&$-$0.21&0.22&$-$0.29&0.22\\
IAF&$-$1.23&0.29&$-$1.26&0.30\\
NCL&\multicolumn{2}{c}{Converged to NM}&\multicolumn{2}{c}{---}\\
IOP&0.52&0.22&\multicolumn{2}{c}{---}\\
CLK&0.52&0.22&\multicolumn{2}{c}{---}\\
\hline
NM$\rm{^{SOC}}$&0&---&\multicolumn{2}{c}{---}\\
FM$\rm{^{SOC}}$&$-$0.31&0.19&\multicolumn{2}{c}{---}\\
IAF$\rm{^{SOC}}$&$-$1.58&0.30&\multicolumn{2}{c}{---}\\
NCL$\rm{^{SOC}}$&\multicolumn{2}{c}{Converged to NM}&\multicolumn{2}{c}{---}\\
IOP$\rm{^{SOC}}$&$-$0.11&0.19&\multicolumn{2}{c}{---}\\
CLK$\rm{^{SOC}}$&$-$0.11&0.19&\multicolumn{2}{c}{---}\\
\end{tabular}
\end{ruledtabular}
\footnotetext[1]{The relaxation was only done for configurations with collinear magnetism because of the bad convergence behavior for those with noncollinear magnetism.}
\footnotetext[2]{The structural optimizations were done with cell parameters fixed.}
\end{table}

Table~\ref{tab:mag} lists the total energies of different initial magnetic configurations. As one can easily read, there is no significant energy difference for all configurations within the DFT error bar of $\sim$~1~meV/atom. More detailed examinations show that SOC has little effect not only on the total energy, but also on the sizes and directions of magnetic moments. Such results indicate that the KCr$_3$As$_3$H should have a paramagnetic ground state, which is similar to the case of K$_2$Cr$_3$As$_3$~\cite{K233.caoc}. However, our results deviate from the result of Taddei \emph{et~al}., in which the total energy for IAF state is $\sim$~5~meV/Cr lower than NM state~\cite{KHCr3As3_NPD}. Such a difference is most likely to be induced by the different setups (e.g., energy cutoff, $\bm{k}$-mesh, or pseudopotentials) we have employed.

\subsection{Electronic Structures}

\begin{figure*}
\centering
\includegraphics[width=18cm]{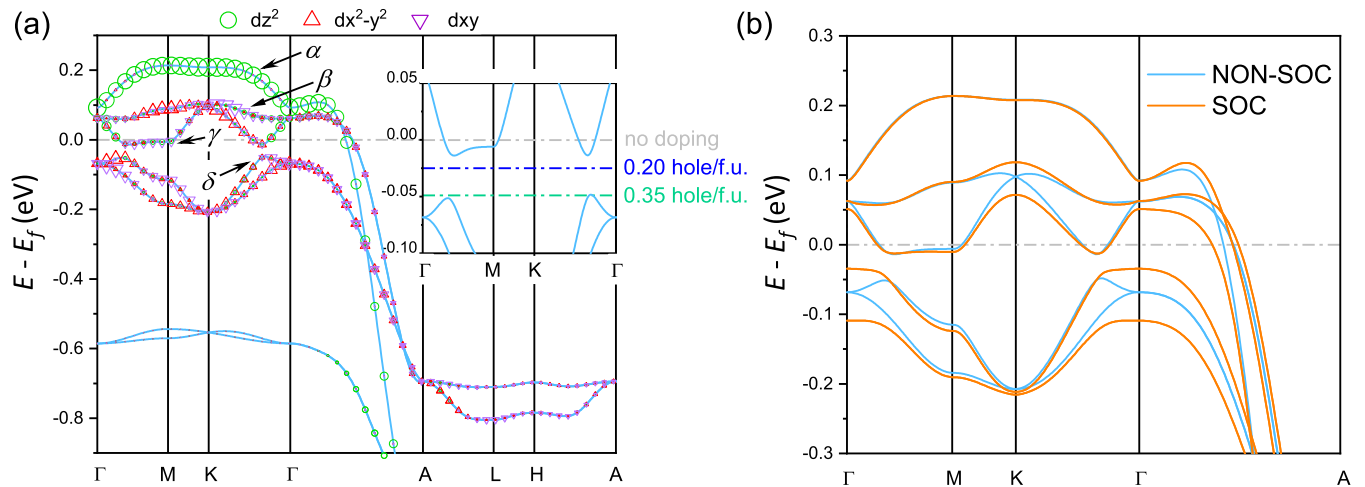}
\caption{\label{Band}Electronic band structures for KCr$_3$As$_3$H. (a) Band structure without SOC effect. The contributions from Cr-d$_{z^2}$, Cr-d$_{x^2-y^2}$ and Cr-d$_{xy}$ orbitals are denoted by the sizes of green circles, red up-triangles, and purple down-triangles, respectively. The inset shows the Fermi-level shifting upon hole doping within rigid band approximation.
(b) SOC effect on electronic band structure. The orange and blue lines represent the band structures with and without SOC respectively.}
\end{figure*}

\begin{figure*}
\centering
\includegraphics[width=18cm]{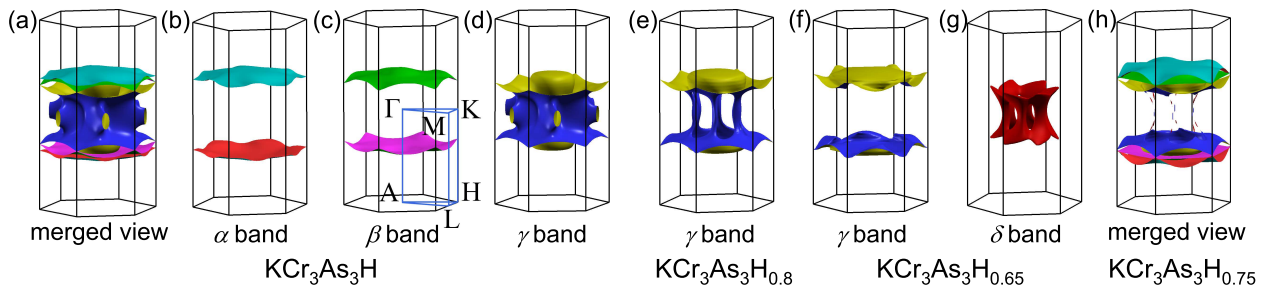}
\caption{\label{FermiSurface}Fermi-surfaces (FS) for KCr$_3$As$_3$H$_x$ of different doping levels. (a-d) Fermi surfaces of (a) merged view, (b) $\alpha$ band, (c) $\beta$ band, and (d) $\gamma$ band for undoped KCr$_3$As$_3$H. (e) FS-$\gamma$ for doping level of 0.2~hole/f.u. (f, g) FS-$\gamma$ and FS-$\delta$ for doping level of 0.35~hole/f.u. (h) merged view of FSs for 0.25~hole/f.u. doped sample. The definitions of high-symmetry lines are illustrated in (c). The doping effects were evaluated within rigid band approximation.}
\end{figure*}

The electronic band structures for KCr$_3$As$_3$H are shown in Fig.~\ref{Band}. Crossing Fermi-level there are three bands, which are mainly contributed by d$_{z^2}$, d$_{x^2-y^2}$ and d$_{xy}$ orbitals of Cr atoms. These bands, marked by $\alpha$, $\beta$, and $\gamma$ in descending order of energy eigenvalues, form three FSs respectively. The $\gamma$ band crosses Fermi-level in the $k_z$~=~0 plane, cuts the $\Gamma$-A line at $k_z$~=~0.21, and forms a 3D FS around the $\Gamma$ point [Fig.~\ref{FermiSurface}(d)]. While the $\alpha$ and $\beta$ bands only cross Fermi-level at around $k_z$~=~0.25. These bands form two Q1D FSs with a little dispersion in $k_xk_y$ plane [Fig.~\ref{FermiSurface}(b, c)], and cross each other at $k_x~=~k_y$~=~0, $k_z$~=~0.23.
Compared with the band structure of KCr$_3$As$_3$ or K$_2$Cr$_3$As$_3$~\cite{K133.caoc,K233.caoc}, The band structure of KCr$_3$As$_3$H holds similar shape to KCr$_3$As$_3$, with only modest distortions and some up-shifting of the Fermi-level. This is not strange since the spatial symmetries are the same for KCr$_3$As$_3$ and KCr$_3$As$_3$H. And with the presence of H, KCr$_3$As$_3$H is effectively doped with electrons. Considering the similar orbital contributions for all the three compounds, the band filling of KCr$_3$As$_3$H is closer to that of K$_2$Cr$_3$As$_3$, which is probably the reason why KCr$_3$As$_3$H has the same paramagnetic ground state as K$_2$Cr$_3$As$_3$ while KCr$_3$As$_3$ holds an IAF one~\cite{K133.caoc}.\\
We then examined the SOC effect on the band structure of KCr$_3$As$_3$H. As shown in Fig.~\ref{Band}(b), after switching on the spin-orbit interaction, the degeneracies at K point and along $\Gamma$-A line are lifted. The SOC splitting at K point is around 50~meV, close to the value in K$_2$Cr$_3$As$_3$. While the effects of SOC splittings are quite different in these two compounds.
In K$_2$Cr$_3$As$_3$, due to the lack of inversion center, the spin-degeneracy of band structure is lifted by the asymmetric SOC (ASOC) effect~\cite{K233.caoc}. The ASOC effect also leads to a sizeable change for the 3D FS sheet.
However, in KCr$_3$As$_3$H, the ASOC effect is ruled out by the presence of spatial inversion symmetry. Therefore in KCr$_3$As$_3$H, the spin-degeneracy of each band persists [Fig.~\ref{Band}(b)], and the FS sheet of $\gamma$ band remains almost unchanged.
The only significant change for Fermi-surfaces is that the two Q1D FSs no longer cross each other at $k_x~=~k_y$~=~0.

By closely checking the band structure along the high-symmetry line $\Gamma$-M-K-$\Gamma$, we find that there is a gap slightly below the Fermi-level. Further examinations of eigenvalues show that the gap remains unclosed over the whole $k_z$~=~0 plane. Considering the presence of H deficiencies in experiments, which would induce effective hole doping for KCr$_3$As$_3$H~\cite{KHCr3As3_NPD},
it is necessary to investigate what would happen if we tune Fermi-level into the gap with hole doping.
As shown in the inset of Fig.~\ref{Band}(a), within rigid band approximation, a modest doping of 0.2~hole/f.u. (nominally KCr$_3$As$_3$H$_{0.8}$) would lower the Fermi-level into the gap. Here one may expect the 3D FS sheet to disappear, but when we examine the FS plot for KCr$_3$As$_3$H$_{0.8}$ [Fig.~\ref{FermiSurface}(e)], we find that the 3D FS sheet doesn't disappear but forms several tube-like connections between the drum-like FS sheets.
If we further tune the doping level to 0.35~hole/f.u., the FS of $\gamma$ band eventually reduce to Q1D [Fig.~\ref{FermiSurface}(f)]. Meanwhile, a new 3D FS sheet of $\delta$ band emerges [Fig.~\ref{FermiSurface}(g)]. The system exhibits a Lifshitz transition in the doping process~\cite{Lifshitz}.
Interestingly, at the transition point (doping level of 0.25~hole/f.u.), only few metallic states are presented in the region between the Q1D FS sheets. Such a reduction of 3D FS sheet indicates a possible enhancement of the Q1D property for KCr$_3$As$_3$H$_{0.75}$, which might be detected by NMR or anisotropic transparent measurements.

\begin{figure}
\includegraphics[width=8cm]{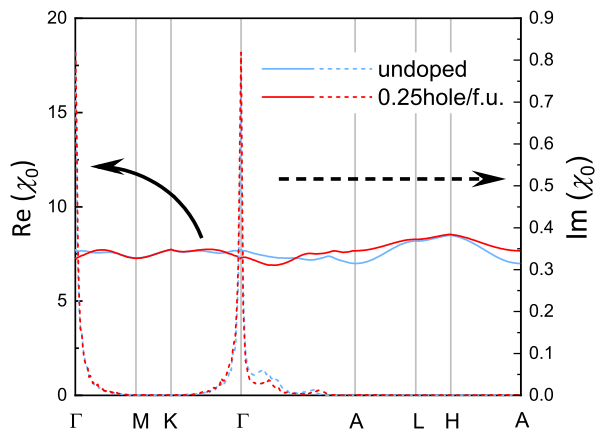}
\caption{\label{Fig:Chi0}Bare electron susceptibilities for KCr$_3$As$_3$H$_x$. The blue lines denote the susceptibility for undoped KCr$_3$As$_3$H. The red lines denote the susceptibility for hole doped KCr$_3$As$_3$H$_{0.75}$.}
\end{figure}

We have also calculated the bare electron susceptibilities $\chi_0$ at $\omega\rightarrow$~0. As illustrated in Fig.~\ref{Fig:Chi0}, the real part and the imaginary part of $\chi_0$ show quite different behaviors. For the real part, the Re($\chi_0$) shows a flat spectrum with no peaks presented. While for the imaginary part, the Im($\chi_0$) exhibits a significant peak at $\Gamma$ point.
The flat feature of Re($\chi_0$) shows a good consistency with the paramagnetic ground state in KCr$_3$As$_3$H, significantly differing from the case in KCr$_3$As$_3$~\cite{133Swave}. The $\Gamma$ peak of Im($\chi_0$) indicates a possible FM spin-fluctuation channel in KCr$_3$As$_3$H$_x$, which may serve as pairing mechanism for the possible spin-triplet superconductivity~\cite{NaCoO2}.

\section{Discussions}
From the electronic band structures of KCr$_3$As$_3$H and KCr$_3$As$_3$, we can see that the presence of H leads to an up-shifting of the Fermi-level in KCr$_3$As$_3$H. Here one may intuitively conclude that the H has metallic bondings with surrounding Cr atoms and acts as an electron donor as K, which seems to be plausible at the first glance. However, when we closely examine the charge distribution in these two compounds, the metallic electron donor picture is violated.

\begin{table}[h]
\caption{\label{tab:Bader}Bader valence charges for different atoms in KCr$_3$As$_3$H and KCr$_3$As$_3$. The values of electrically neutral atoms are also listed here for comparison.}
\begin{ruledtabular}
\begin{tabular}{cccc}
&\multicolumn{3}{c}{Bader Charge}\\
Element&KCr$_3$As$_3$H&KCr$_3$As$_3$&Neutral atom\\
\hline
K&8.22&8.22&9\\
Cr&11.37&11.48&12\\
As&15.73&15.78&15\\
H&1.47&---&1\\
\end{tabular}
\end{ruledtabular}
\end{table}

Table~\ref{tab:Bader} lists the Bader valence charges for different atoms in KCr$_3$As$_3$H and KCr$_3$As$_3$~\cite{Bader.Sanville,Bader.Tang}. Compared with neutral atoms, it is obvious that H and K have different valence states. For K, the valence electrons are well-delocalized from Bader atom region, leaving K a positive chemical valence with fewer Bader charges.
While for H, on the contrary, more electrons are attracted into the Bader atom, which results in a negative chemical valence. From KCr$_3$As$_3$ to KCr$_3$As$_3$H, the decreased charges for Cr and As per unit cell are 0.67~e$^-$/$u.c.$ and 0.27~e$^-$/$u.c.$, which exactly compensates the increased Bader charges of H atoms.
Such results indicate that instead of acting as a metallic electron donor, the H atoms conversely receive electrons from CrAs tubes in KCr$_3$As$_3$H.

\begin{figure}
\includegraphics[width=8cm]{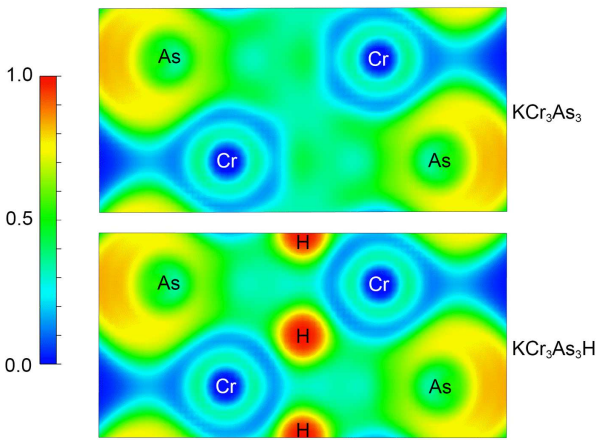}
\caption{\label{ELF}2D view of electron localization functions in KCr$_3$As$_3$ and KCr$_3$As$_3$H. The plane is sliced along the (1$\overline{5}$0) surface and crossing the original point (shown as the red dashed line in Fig.~\ref{Struct}(b)).}
\end{figure}

The non-metallic feature of H is further confirmed by our calculations of electron localization function (ELF)~\cite{ELF.1990}, which is defined by:
\begin{equation}
\label{equ:ELF}
{\rm ELF}(\bm{r})~=~\frac{1}{1+(D_{\sigma}(\bm{r})/D^0_{\sigma}(\bm{r}))^2}
\end{equation}
Where $D_{\sigma}(\bm{r})$~=~$\sum\limits_{i}\left|\nabla\psi_{i\sigma}(\bm{r})\right|^2$~-~$\frac{1}{4}\frac{(\nabla\rho_{\sigma}(\bm{r}))^2}{\rho_{\sigma}(\bm{r})}$, is proportional to the probability of finding another electron of spin $\sigma$ around the ``detective'' spin-$\sigma$ electron at $\bm{r}$. And $D^0_{\sigma}(\bm{r})$~=~$\frac{3}{5}(6\pi^2)^{2/3}(\rho_{\sigma}(\bm{r}))^{5/3}$ denotes the value for the uniform electron gas. Here in DFT, the $\psi_{i\sigma}(\bm{r})$ and $\rho_{\sigma}(\bm{r})$ represent the Kohn-Sham wave function and charge density respectively. The index $i$ goes through all the occupied states in the summation of $D_{\sigma}(\bm{r})$.\\
\noindent
By the construction of ELF, we can see that its value falls between 0 and 1. And since the stronger localization corresponds to the smaller value of $D_{\sigma}$, the value of ELF gets closer to 1 as the electron gets more localized. Specially, if the value of ELF equals 0.5, it yields $D_{\sigma}$~=~$D^0_{\sigma}$, such a value reflects a localization level similar to that of the uniform electron gas.\\
\noindent
Our results for ELF calculations are shown in Fig.~\ref{ELF}, which exhibit distinct behaviors from the picture of metallic donor.
For a metallic donor, the electrons around it should be delocalized, and forms attractors (regions where ELF shows local maximum) in the interstitial region~\cite{ELF.nature}. While for the results of our ELF calculations, we can immediately see that the electrons around H are highly localized. And if we check the interstitial region, it exhibits a uniform electron gas-like behavior with no attractor presented. Thus in KCr$_3$As$_3$H, the H atom shouldn't serve as a simple electron donor with metallic bondings. Instead, it acts as a strong electron attractor. The strong-attractor behavior is also consistent with our previous Bader charge analysis.

\begin{figure*}
\includegraphics[width=16cm]{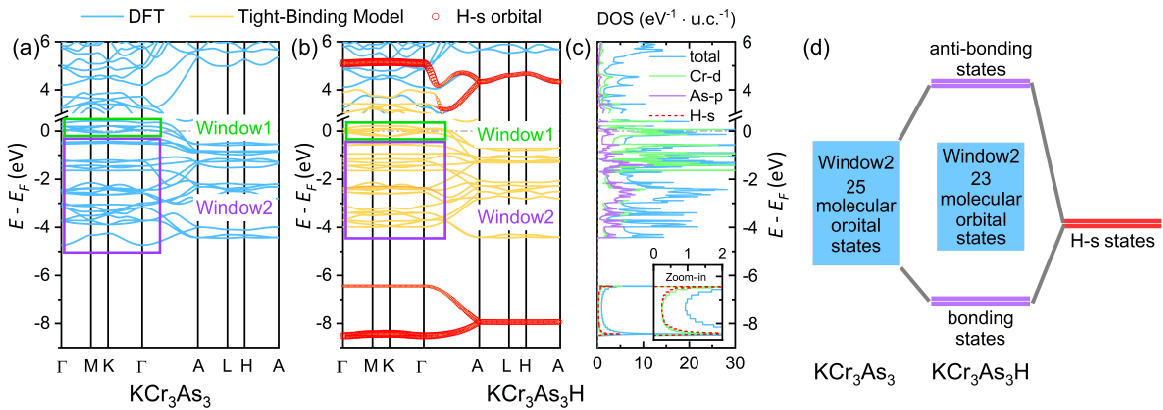}
\caption{\label{H-bond}Electronic structures for KCr$_3$As$_3$ and KCr$_3$As$_3$H with a larger energy scale. (a) Band structure of KCr$_3$As$_3$. (b) Band structures from DFT (blue lines) and tight-binding model (orange lines) for KCr$_3$As$_3$H. The sizes of red circles are proportional to the contributions of H-s orbitals in our tight-binding model. (c) Projected density of states for KCr$_3$As$_3$H obtained from DFT calculations. The inset shows enlarged view for the bonding states.} (d) Sketch of the bonding mechanism between H-s and molecular orbital states in KCr$_3$As$_3$H.
\end{figure*}

Now here comes a question: Since H exhibits negative valence state in KCr$_3$As$_3$H, it should attract more electrons from conduction bands and results in \emph{hole} doping. But from our previous calculations, the presence of H in KCr$_3$As$_3$H turns out to induce an effective \emph{electron} doping. So, how is the effective electron doping induced?\\
\noindent
To understand this question, we investigated the electronic structures for KCr$_3$As$_3$ and KCr$_3$As$_3$H with a larger energy scale.
As shown in Fig.~\ref{H-bond}(a, b), the band structures around Fermi-level (window-1) are quite similar for KCr$_3$As$_3$ and KCr$_3$As$_3$H. While at around \mbox{$E$~-~$E_F$~=~$-$8~eV}, there emerge two new bands with little $k_xk_y$ dispersion for KCr$_3$As$_3$H. The projected density of states (PDOS) [Fig.~\ref{H-bond}(c)] shows that these bands are mainly contributed by the bondings between H-s and Cr-d orbitals. And when we check the projection of H-s orbitals from our tight-binding model, another two anti-bonding states can be found well above Fermi-level [Fig.~\ref{H-bond}(b)]. Close examinations of window-2 show that the band numbers in window-2 are 25 for KCr$_3$As$_3$ and 23 for KCr$_3$As$_3$H. The two missing bands in KCr$_3$As$_3$H exactly corresponds to the states that form bondings with H.
To specify the orbital origination of the bonding states, we introduce Cr-triangular centered molecular orbitals, which correspond to two sets of 2D $E$ representation and one 1D $A_1'$ representation of the $D_{3h}$ point group~\cite{djh}. Due to the $D_{3h}$ symmetry of the crystal field, only the hopping terms between H-s and molecular-$A_1'$ orbitals could be non-zero. Thus in KCr$_3$As$_3$H, the H atoms only form bondings with molecular-$A_1'$ orbitals.\\
\noindent
The bonding picture of H atoms then offers us a possible explanation for the effective electron doping:
As shown in Fig.~\ref{H-bond}(d), for each unit cell in KCr$_3$As$_3$H, the H atoms form two pairs bonding and anti-bonding states with molecular-$A_1'$ orbital states in window-2. And since the two anti-bonding states are pushed well above Fermi-level, the presence of H doesn't introduce any additional unoccupied states below Fermi-level. Therefore, from KCr$_3$As$_3$ to KCr$_3$As$_3$H, the net effect of H is to introduce 2 additional electrons/u.c., which eventually induces the nontrivial electron doping in KCr$_3$As$_3$H.

\section{Summary and Conclusions}

In summary, we have studied the magnetic ground state, electronic structure, and bare electron susceptibility for KCr$_3$As$_3$H with DFT calculations. The total energy calculations show that KCr$_3$As$_3$H has a paramagnetic ground state. The band structure of KCr$_3$As$_3$H exhibits one 3D Fermi-surface and two Q1D Fermi-surfaces which are mainly contributed by d$_{z^2}$, d$_{x^2-y^2}$ and d$_{xy}$ orbitals of Cr atoms. Upon moderate hole doping, the system undergoes a Lifshitz transition. The Q1D feature of the system might be enhanced at the transition point. The bare electron susceptibility calculation indicates possible ferromagnetic spin fluctuations in KCr$_3$As$_3$H, which may serve as the pairing mechanism for the possible spin-triplet superconductivity.\\
With the tools of Bader charge analysis and electron localization functions, we also examined the bonding features of hydrogen in KCr$_3$As$_3$H. Our results show that the H atom acts as an electron acceptor, and strongly bonds with the surrounding Cr atoms.
Therefore the hydrogen doping helps to eliminate the imaginary phonon frequencies in KCr$_3$As$_3$~\cite{KHCr3As3_NPD, 133.singh} by forming strong bonding with Cr atoms and stabilizing the CrAs-tube.
It also takes the responsibility for the effective electron doping in KCr$_3$As$_3$H.
Our strong-bonding picture of H is also in good consistence with the high H vibrational frequency of $\sim$~35~THz and the large formation energy of $-$92~kJ/mol~H$_2$ for KCr$_3$As$_3$H~\cite{KHCr3As3_NPD}.

\begin{acknowledgments}
The authors would like to thank Chen-Chao Xu and Guo-Xiang Zhi for instructive discussions.
This work was supported by the National Natural Science Foundation of China (NSFC, No.~11674281 and No.~11874137). All the calculations were performed at the National Supercomputer Center in Guangzhou.
\end{acknowledgments}

\bibliography{KCr3As3H}

\end{document}